\documentclass[10pt,journal,compsoc]{IEEEtran}
\ifCLASSOPTIONcompsoc
  \usepackage[nocompress]{cite}
\else
  \usepackage{cite}
\fi

\ifCLASSINFOpdf
  \usepackage[pdftex]{graphicx}
\else
\fi

\usepackage{amsmath}

\usepackage{multirow,tabularx}

\usepackage{caption}

\newcommand\MYhyperrefoptions{bookmarks=true,bookmarksnumbered=true,
pdfpagemode={UseOutlines},plainpages=false,pdfpagelabels=true,
colorlinks=true,linkcolor={black},citecolor={black},urlcolor={black},
pdftitle={A Fast Hardware Pseudorandom Number Generator Based on xoroshiro128},
pdfsubject={Hardware Pseudorandom Number Generator Design},
pdfauthor={James Hanlon, Diya Rajan, Stephen Felix},
pdfkeywords={Pseudorandom Number Generator, PRNG, Statistical Tests}}

\ifCLASSINFOpdf
\usepackage[\MYhyperrefoptions,pdftex]{hyperref}
\else
\usepackage[\MYhyperrefoptions,breaklinks=true,dvips]{hyperref}
\usepackage{breakurl}
\fi

\begin{document}

\title{A Fast Hardware Pseudorandom Number Generator Based on xoroshiro128}

\author{James~Hanlon and~Stephen~Felix%
\thanks{The authors are from Graphcore Ltd., Bristol, BS1 2PH, UK.\protect\\
E-mail: \{jamie.hanlon,stephen.felix\}@graphcore.ai}%
\thanks{Manuscript received January 18th, 2022; revised 3rd August 2022.}%
\thanks{\copyright 2022 IEEE.  Personal use of this material is permitted.
Permission from IEEE must be obtained for all other uses, in any current or
future media, including reprinting/republishing this material for advertising
or promotional purposes, creating new collective works, for resale or
redistribution to servers or lists, or reuse of any copyrighted component of
this work in other works.}}

\newcommand\ftwo{$\boldsymbol{{\rm F}}_2$}


\IEEEtitleabstractindextext{%
\begin{abstract}

The Graphcore Intelligence Processing Unit contains an original pseudorandom
  number generator (PRNG) called xoroshiro128aox, based on the \ftwo-linear
  generator xoroshiro128. It is designed to be cheap to implement in hardware
  and provide high-quality statistical randomness.
In this paper, we present a rigorous assessment of the generator's quality
  using standard statistical test suites and compare the results with the fast
  contemporary PRNGs xoroshiro128+, pcg64 and philox4x32-10.
%
We show that xoroshiro128aox mitigates the known weakness in the lower order
  bits of xoroshiro128+ with a new 'AOX' output function by passing the
  BigCrush and PractRand suites, but we note that the function has some minor
  non uniformities. 
We focus our testing with specific tests for linear artefacts to highlight the
  weaknesses of both xoroshiro128 PRNGs, but conclude that they are hard to
  detect, and xoroshiro128aox otherwise provides a good trade off between
  statistical quality and hardware implementation cost.


\end{abstract}

\begin{IEEEkeywords}
Pseudorandom Number Generator, PRNG, Hardware Circuits, Statistical Tests
\end{IEEEkeywords}

}

\maketitle

\IEEEdisplaynontitleabstractindextext

%
\IEEEpeerreviewmaketitle

\ifCLASSOPTIONcompsoc
\IEEEraisesectionheading{\section{Introduction}\label{sec:introduction}}
\else
\section{Introduction}
\label{sec:introduction}
\fi

%
%
%
%

Randomness is widely used in machine intelligence (MI) algorithms.
Examples include shuffling of data prior to each training epoch for
stochastic gradient descent~\cite{Saad98}, sub sampling of training images,
weight initialisation, adding noise to activations or weights, regularisation
techniques like Dropout~\cite{Srivastava14} and Stochastic
Pooling~\cite{Zeiler13}, Monte Carlo sampling in generative models or choosing
random actions in reinforcement-learning models.
%
The use of pseudorandom number generators (PRNGs), deterministic
algorithms for generating sequences of numbers that appear to be random, are
ubiquitous in computing. When compared with true random number generators
(TRNGs) that are typically based on sampling of some physical phenomena such as
ring oscillators~\cite{Varchola10}, PRNGs offer a higher rate of output, no
requirement for special hardware structures, and unlike TRNGs, have the ability to be
\emph{seeded} to replay a sequence deterministically.
In application domains such as AI, there is no need for a PRNG to be
cryptographically secure, meaning that it does not need to be difficult for an
adversary to predict future outputs based on past ones.

It is not clear to what degree the statistical quality of a PRNG affects the
performance of MI applications.
For example, a minor correlation between the initial values of weights in a
neural network may have a negligible impact since backpropogation tunes weight
values and should lead to locally optimum solutions regardless of whether the
weights were perfectly randomly distributed initially. On the other hand,
applications such as Monte-Carlo approximation can produce inaccurate solutions
when based on correlated PRNG output~\cite{Ferrenberg92}.
It is interesting to note that Python’s default PRNG is the 32-bit Mersenne
Twister algorithm,\footnote{See Python 3's random module documentation:
\url{https://docs.python.org/3/library/random.html}} which fails standard
statistical tests for linearity~\cite{Vigna19}.
Failures for specific statistical correlations can be important or anecdotal.
Linear correlations, which are a main focus of this paper, do not matter if
their effects are diluted by way in which they are used, such as to create
uniform floating point values manipulated by non-linear arithmetic operations.
Other beneficial aspects of a PRNG, such as performance, memory requirements or
hardware implementation cost, must be weighed against particular statistical
weaknesses.

Software PRNGs can be implemented with a low overhead of just a few
instructions per output. However, when randomness is required more frequently,
generation in hardware can provide performance that is orders of magnitude
better than compatible generation in software.
In the Graphcore Intelligence Processing Unit (IPU)~\cite{Knowles21}, each of
its 1,216 tile processors contains a novel PRNG called \texttt{xoroshiro128aox}
that is capable of producing 64 bits of random data every cycle. This
randomness is used either to automatically round floating-point numbers
stochastially~\cite{GraphcoreAIFloat} or is made available to the programmer
through instructions to generate random values in uniform and Gaussian
distributions.

This paper presents \texttt{xoroshiro128aox} and a rigorous assessment of its
quality using standard statistical tests. Our results indicate that the PRNG
is comparable to contemporary fast non-cryptographic PRNGs with similar
state size, whilst being cheaper to implement in hardware.
The remainder of this paper is structured as follows.
Section~\ref{statistical-testing} describes standard PRNG statistical testing,
and the specific test sets that are used in our investigation.
Section~\ref{xoroshiro128aox} introduces the IPU's PRNG and the
\texttt{xoroshiro128} family that it is derived from.
Section~\ref{methodology} describes the methodology used to perform the
statistical testing, as well as the other generators included for comparison.
Section~\ref{results} presents the results of the empirical statistical
analysis.
Section~\ref{implementation-cost} analyses the hardware implementation cost of
the generators considered by synthesising them in hardware.
Section~\ref{other-quality} identifies several other aspects of PRNG quality
and analyses these for \texttt{xoroshiro128aox}.
Section~\ref{conclusion} concludes the investigation.

\section{Statistical testing}
\label{statistical-testing}

Theoretical analysis of a PRNG can be used to establish some properties, such
as values being produced uniformly and over their entire period length,
however only empirical testing can be used to establish the statistical
properties of a PRNG, and is the standard approach for judging quality.
An empirical statistical test involves sampling the output of a generator,
calculating a summary statistic such as mean or standard deviation,
then comparing this to the same statistic for a truly random source. This
approach is only applicable when the number of samples is less than the
sequence length, since it would otherwise be easy to detect a repeating
sequence.

Empirical testing of PRNGs is formalised by comparing against a null
hypothesis, where we assume the output of a generator follows a uniform
distribution as would be the case for a true random generator. A particular
test calculates a statistic that has a known distribution under the null
hypothesis using on a finite portion of the generator's output. Probabilities
called $p$-values are calculated based on how likely it is that the generator's
output is consistent with the test statistic's distribution. Extreme $p$-values
that are very close to 0 or 1 indicate that the sampled output is unlikely to
be random.
In statistical testing of PRNGs, results are categorised such that extreme
$p$-values will be flagged as 'suspicious' or 'anomalous', but to determine a
pass or fail result, an arbitrary threshold can be applied. The exact bounds
may depend on the methodology.

Because an unlimited number of statistical tests can be devised and each test
will explore different aspects of the PRNG, it is not possible for any set of
tests to determine whether a generator is perfectly random. Indeed, since we
are interested in designing a generator that is fast and cheap to implement,
we are concerned only that the generator is good: that it passes all simple
tests and fails complex tests infrequently~\cite{Ecuyer13}.
For our empirical analysis, we use the statistical test suites for RNGs
provided by the TestU01~\cite{Ecuyer07}, PractRand~\cite{Humphrey} and
Gjrand~\cite{Blackman} libraries. These are all well regarded by the community
for their ability to distinguish good from bad PRNGs. Notable other test suites
are DieHarder, RaBiGeTe and NIST STS, but are less comprehensive, less well
regarded by the community and less well maintained.

TestU01 provides implementations of standard statistical tests such as Birthday
Spacings that measures the distribution of distances between outputs and
Collision that measures the probability of outputs occurring in the same
interval of bits multiple times.
Sets of statistical tests are instanced in test batteries that can be used for
exercising RNGs. The most stringent battery BigCrush contains 106 test
parameterisations of 30 individual tests, and consumes close to 1~TB of random
data.\footnote{The tests conducted and the parameters used are explained in
detail in the TestU01 User Guide~\cite{Ecuyer13}.}
Each test can produce one or more $p$-values, and in total a single run of
BigCrush produces 160 $p$-values. In total, 254 $p$-values are reported in the
output of BigCrush, however, only independent $p$-values are reported in the
summary, which we count towards failures in our analysis.

PractRand (Practically Random) provides statistical testing of RNGs using six
tests,\footnote{See the PractRand documentation for details of these tests
\url{http://pracrand.sourceforge.net/Tests_engines.txt}} deployed in many
different parameterisations. Several of these tests are original and developed
by the software's author Chris Doty-Humphrey, making it a complement to
BigCrush. PractRand provides the ability for generators to be tested with
effectively unlimited sequence lengths, although the default sequence length is
32~TB (requiring approximately a week of run time), which we use in our
analysis. For this test length, PractRand runs 455 test parameterisations of
the six individual tests.

Gjrand is another library of PRNGs and statistical tests, created by David
Blackman, a co-creator of the \texttt{xoroshiro128} family of generators. Its
test suite for uniform random bits performs 13 tests and is limited to a
maximum of 10~TB of output, which we use for our analysis.

\section{xoroshiro128aox}
\label{xoroshiro128aox}

The IPU's PRNG is based on \texttt{xoroshiro128}, an \ftwo-linear engine
developed by Blackman and Vigna in 2017~\cite{Blackman21}. It operates by
performing the operations exclusive or (XOR), rotate, shift and rotate
consecutively on 128 bits of state.
Any generator based on an \ftwo-linear map will fail tests such as binary rank
and linear complexity, which are designed to detect linear artefacts.
A family of more robust \texttt{xoroshiro128} PRNGs is obtained by adding a
non-linear function of the state vector to 'scramble' the output. Use of such
an output function reduces or eliminates linear artefacts, improving the
statistical properties of the generator. Blackman and Vigna suggest the
scrambling functions: \texttt{+} (addition), \texttt{*} (multiplication),
\texttt{++} (sum, rotation, sum) and \texttt{**} (multiplication, rotation,
multiplication), which are cheap to execute in modern processors.

\texttt{xoroshiro128+} has received particular attention because it is the
fastest variant of this family and its predecessor \texttt{xorshift128+} is
used in the JavaScript engines of Chrome,\footnote{See this 2015 blog post from
the Google V8 project: \url{https://v8.dev/blog/math-random}.} Firefox and
Safari web browsers. However, the use of addition as a non-linearity leaves the
least significant bits as a weak linear combination of the state vectors, and
in particular that bit 0 is just the XOR of its two input bits.
Blackman and Vigna acknowledge this weakness by showing that the linearities
are detectable by the \texttt{MatrixRank} and \texttt{LinearComp} tests of
TestU01 only when the least significant bits of the output are placed in the
most significant positions. 
This is because TestU01's Crush batteries are designed to test random
floating-point numbers in the range $[0,1)$, conversion of random bits biases
the higher bits since the lowest bits will be most affected by numerical
errors.

In our own testing, which is detailed in later sections, we observe the
weakness of the lower bits when providing the bit-reversed 32-bit output to
TestU01, as well as in other tests. Similar results have been reported by
Lemire and O'Neill across various PRNGs using addition as an output non
linearity~\cite{Lemire19}.

%


The Graphcore IPU's \texttt{xoroshiro128aox} generator uses the
\texttt{xoroshiro128} PRNG with a non-linear operation based on a sequence of
AND, OR and XOR operations. This new output function has been designed to hide
linearities and to be cheap to implement in hardware, particularly compared
with 64-bit addition.
Every output bit of AOX is dependent on the same number of input bits according
to a similar pattern, with dependencies on bits to the left and right (more and
less significant), compared with addition where dependencies are only on less
significant right-hand bits.
A disadvantage of this feature of AOX is that it is not possible
to prove the output is uniform, which we analyse in Section~\ref{uniformity},
whereas addition is provably uniform.
From two 64-bit state vectors $s0=\{s0_0,s0_1,\cdots, s0_{63}\}$ and
$s1=\{s1_0,s1_1,\cdots, s1_{63}\}$, result bit $i$ of the output $r$ is defined
as:
\begin{equation}
\begin{split}
  r_i = s0_i \oplus s1_i \oplus (&(s0_{(i-1)\bmod{64}} \wedge s1_{(i-1)\bmod{64}}) \vee \\
   & (s0_{(i-2)\bmod{64}} \wedge s1_{(i-2)\bmod{64}}))
\end{split}
\end{equation}

Figure~\ref{aox-c} lists a C implementation of the \texttt{xoroshiro128aox}
generator, with a function \texttt{next} that advances the 128-bit state and
returns 64 random bits.
%
The shift constants 55, 14 and 36 are from the 2016 version of
\texttt{xoroshiro128+},\footnote{This is noted in the \texttt{xoroshiro128+}
source code \url{https://prng.di.unimi.it/xoroshiro128plus.c}} which we
used in our 2017 IPU silicon implementation. Blackman and Vigna later proposed
24, 16 and 37 as producing superior output. We include results in this
investigation for both variants of the constants to show there are no
significant statistical differences.

\begin{figure}[!t]
%
%
%
\begin{verbatim}
uint64_t s0, s1; // State vectors

uint64_t rotl(uint64_t x, int k) {
  return (x << k) | (x >> (64 - k));
}

uint64_t next(void) {
  uint64_t sx = s0 ^ s1;
  // Calculate the result, the 'AOX' step.
  uint64_t sa = s0 & s1;
  uint64_t res =
    sx ^ (rotl(sa, 1) | rotl(sa, 2));
  // State update
  s0 = rotl(s0, 55) ^ sx ^ (sx << 14);
  s1 = rotl(sx, 36);
  return res;
}
\end{verbatim}
\caption{A C implementation of \texttt{xoroshiro128aox}.
For \texttt{xoroshiro128+}, \texttt{res} is calculated as \texttt{s0 + s1}.}
\label{aox-c}
\end{figure}



\section{Generators for comparison}

To provide a baseline result, tests are performed against
\texttt{xoroshiro128+} and with two fast contemporary and comparable PRNGs,
both with 128 bits of state and 64 bits of output: \texttt{philox4x32-10} and
\texttt{pcg64}.

Philox~\cite{Salmon11} is a counter-based family of PRNGs that have a simple
state transition function of an increment by one, but a complex output function
to map state and key values to pseudorandom outputs. The state transition makes
it easy to jump to arbitrary points in the sequence by just setting the
counter, which is useful for initialising parallel generators.
We choose the \texttt{philox4x32-10} variant of this family which has a 128-bit
integer counter and two 32-bit keys as its internal state. Although this makes
the complete state size 192 bits, this is the most closely comparable version
of Philox to \texttt{xoroshiro128+}.
The output of \texttt{philox4x32-10} is calculated by performing ten rounds of
a scrambling function composed of 32-bit multiplications and 32-bit XORs. The key
values, which are used as inputs to this are incremented each round by a
constant value.
%
%
The \texttt{philox4x32-10} generator has established itself as a standard, and
consequently is available as part of Python scientific computing library NumPy
and Nvidia's GPU cuRAND library.\footnote{See details in
\url{https://developer.nvidia.com/curand}}


\texttt{pcg64} is a linear congruential generator (LCG), which uses
multiplication and addition by constants for the state transition
function~\cite{ONeill14}. To produce outputs, it uses XOR and rotation
operations, in particular using part of the state vector to set a variable
rotation distance.
The generator is specifically characterised by an 'XSL RR' output function,
meaning 'fixed XOR shift to low bits and random rotate', and is part of the PCG
family of PRNGs that are claimed to be fast and high quality compared with
contemporary generators.
%


Finally, the 32-bit Mersenne Twister~\cite{Matsumoto98} (referred to as
\texttt{mt32}) is included in our analysis since it is the most widely used PRNG
in software, and included as the default generator in many software systems
including Microsoft Excel, Python and MATLAB. This generator has a state size
of 2.5~KB (624 32-bit words, or 19,937 bits) and a huge period of $2^{19937} - 1$.
%

The PRNGs used for comparison are tested using reference C/C++ implementations
provided by their authors, or as part of standard libraries. Additional standard
PRNGs are not included in our analysis due to the computational cost of
performing the statistical tests and because results for their quality may
readily be found in the literature.

\section{Methodology}
\label{methodology}

We adopt Vigna's methodology of \emph{sampling generators} for conducting our
tests with BigCrush, PractRand and Gjrand~\cite{Vigna16}, as originally
suggested in~\cite{Ecuyer07}. A generator is tested against a particular test
or test suite suite by choosing 100 seeds spaced equidistantly in the $n$-bit
natural number sequence, that is at intervals $1+i\lfloor2^n/100\rfloor$ for $0
\leq i < 100$, and obtaining results for all 100 seeds. In effect, the seeds are
chosen randomly with respect to the sequence produced by a particular
generator.
It would be preferable to choose seeds spaced equidistantly in a generator's
sequence, but it is not always possible for a generator to jump to arbitrary
points, so this method takes the simplest and most general approach.

For each seed, a generator fails that seed if an extreme $p$-value is reported.
We choose the range of extreme $p$-values to be outside of $[0.001, 0.999]$
across all tests run, which is the default used by TestU01 for reporting
failures. At a particular threshold, a certain number of failures are always
expected: assuming all tests contributing to the score are independent, the
probability that a true random number generator produces a $p$-value outside of
this range is 0.2\%. This probability reduces as the acceptable $p$-value range
increases, as does the required generator output to establish a failing $p$-value.
The criteria for distinguishing a failure is more stringent: a generator fails
a test \emph{systematically} if it fails all seeds on the same test. Where
systematic failures occur, we report the test that caused the failure.
Only the generators that have a systematic failure are considered to fail the
particular test set.

Since AOX is designed to hide the linearities of \texttt{xoroshiro128}, we
apply a more stringent analysis using specific tests that are sensitive to
linearities.
These are the Binary Rank and Linear Complexity tests from TestU01, and a
Hamming-Weight Dependency (HWD) test that is a development of the \texttt{z9}
test included in Gjrand.\footnote{This relationship is mentioned in the source
code for the HWD test: "the Hamming-weight dependency test based on z9 from
gjrand 4.2.1.0", \url{http://xoshiro.di.unimi.it/hwd.c}.}~\cite{Vigna22}
The Matrix Rank test fills a matrix with random values and then computes the
rank (the number of linearly-independent rows), comparing this against the
expected distribution for a random matrix.\footnote{Matrix Rank also appears in
PractRand as \texttt{BRank} and Gjrand as \texttt{binr}.}
The Linear Complexity test is based on output from the Berlekamp–Massey
algorithm that can determine a minimal polynomial for a sequence that is
linearly recurrent, as is the case for an \ftwo-linear map.
The HWD test counts set bits and analyses dependencies between the numbers of
ones and zeros in consecutive outputs, which are correlations indicative of the
sparse matrix representations of the \ftwo-linear map.\footnote{TestU01
provides similar tests that count the frequency of set bits:
\texttt{HammingWeights2}, \texttt{HammingIndep} and \texttt{HammingCorr} and
PractRand with \texttt{DC6}, and \texttt{BCFN}.}

\section{Results}
\label{results}

\subsection{TestU01's BigCrush}

In total, we use six different permutations of the 64-bit PRNG output as input
to the BigCrush test suite to avoid biasing certain bits and to expose known
failures.
Following standard methodology to avoid biasing of the higher-order bits on
conversion to floating-point values in the range $[0,1)$, the standard output
of the generator as well as the bit reverse is taken (referred to as \texttt{std32}
and \texttt{rev32} respectively).
To demonstrate the systematic failures exhibited by \texttt{xoroshiro128+} and
that this same weakness does not exist in other generators, the bit reversal of
the lowest 32 bits are taken as output (referred to as \texttt{rev32lo}).
We remark that this particular manipulation of the output of the generator is
not the only way to expose the weak lower bits to cause a systematic failure.
Experimentally, we have found that it is possible to do so by permuting the
complete 64-bit output firstly by swapping the high and low 16 bits of each
32-bit output and by a particular interleaving of low and high bits over the
full 64-bit output.
For completeness, \texttt{std32lo}, \texttt{std232hi} and \texttt{rev32hi}
output bit permutations are also included, and are summarised in
Table~\ref{output-bits}.

\begin{table}
\renewcommand{\arraystretch}{1.3}
\centering
\caption{Summary of the bits provided to TestU01's BigCrush for each generator}
\begin{tabular}{lll}
\hline
Output & Bits output & Comment \\
\hline
\texttt{std32}   & \texttt{[31:0]}, \texttt{[63:32]} & All 64 bits used \\
\texttt{rev32}   & \texttt{[0:31]}, \texttt{[32:63]} & All 64 bits used \\
\texttt{std32lo} & \texttt{[31:0]}                   & Upper 32 bits discarded \\
\texttt{rev32lo} & \texttt{[0:31]}                   & Upper 32 bits discarded \\
\texttt{std32hi} & \texttt{[63:32]}                  & Lower 32 bits discarded \\
\texttt{rev32hi} & \texttt{[32:63]}                  & Lower 32 bits discarded \\
\hline
\end{tabular}
\label{output-bits}
\end{table}

Table \ref{big-crush-test-failures-summary} provides a summary of BigCrush test
failures for each generator. The number of test failures for a particular
generator and output is the total number of test failures across all 100 seeds.
The \texttt{mt32} generator exhibits a systematic failure for the
\texttt{LinearComp} test across all output permutations. As expected,
\texttt{xoroshiro128+} exhibits a systematic failure for the
\texttt{LinearComp} and \texttt{MatrixRank} tests when run with the
\texttt{rev32lo} output.
The remaining generators do not exhibit any systematic failures and all can be
considered to pass BigCrush.
Note that the number of failures for generators that do not fail systematically, fall
within three standard deviations of the expected value, 32.

\begin{table*}[!t]
\renewcommand{\arraystretch}{1.3}
\centering
\caption{Summary of BigCrush test failures for each generator}
\begin{tabular}{lcccccccl}
\hline
Generator & \texttt{s32} & \texttt{r32} & \texttt{s32l} & \texttt{s32h} & \texttt{r32l} & \texttt{r32h} & Total & Systematic failures \\
\hline
\texttt{mt32}                     & 236 & 237 & 233 &  238 &  246 &  237 &  1427  & \texttt{LinearComp} \\
\texttt{pcg64}                    & 34  & 30  & 38  &  37  &  38  &  27  &  204   & \\
\texttt{philox4x32-10}            & 33  & 32  & 32  &  32  &  28  &  38  &  195   & \\
\texttt{xoroshiro128+-24-16-37}   & 33  & 29  & 28  &  40  &  353 &  42  &  525   & \texttt{LinearComp}, \texttt{MatrixRank} \\
\texttt{xoroshiro128+-55-14-36}   & 31  & 36  & 43  &  39  &  335 &  40  &  524   & \texttt{LinearComp}, \texttt{MatrixRank} \\
\texttt{xoroshiro128aox-24-16-37} & 31  & 32  & 41  &  30  &  44  &  32  &  210   & \\
\texttt{xoroshiro128aox-55-14-36} & 32  & 33  & 44  &  32  &  41  &  31  &  213   & \\
\hline
\end{tabular}
\label{big-crush-test-failures-summary}
\end{table*}

%

\subsection{PractRand}



Table \ref{practrand-results} lists results for the PractRand test set.
Both variants of \texttt{xoroshiro128+} fail quickly as expected, with
systematic failures on three instances of the Binary Rank test. The Mersenne
Twister lasts longer on the Binary Rank tests, but also fails eventually at
256~GB output. The AOX variants show similar results to \texttt{pcg64} and
\texttt{philox4x32-10}, running to completion of the test.

Note that unlike BigCrush, the number of test failures do not follow a
Poisson distribution given because there are high degrees of correlation
between tests and therefore the calculated $p$-values are not entirely
independent, given that there are only six basic tests and thousands of
permutations of their parameters.


\begin{table*}[!t]
\renewcommand{\arraystretch}{1.3}
\centering
\caption{Results for the PractRand test set}
\begin{tabular}{lcccp{0.38\linewidth}}
\hline
Generator & Total failures & Total tests & Total output & Systematic failures \\
\hline
\texttt{mt32}                     & 151 & 36900 & 256 GB & \texttt{[Low16/64]BRank(12):12K(1)} \\
\texttt{pcg64}                    & 72  & 45500 & 32 TB  & \\
\texttt{philox4x32-10}            & 52  & 45500 & 32 TB  & \\
\texttt{xoroshiro128+-24-16-37}   & 401 & 21300 & 256 MB & \texttt{[Low4/64]BRank(12):768(1)}, \texttt{[Low1/64]BRank(12):256(2)}, \texttt{[Low1/64]BRank(12):384(1)} \\
\texttt{xoroshiro128+-55-14-36}   & 402 & 21300 & 256 MB & \texttt{[Low4/64]BRank(12):768(1)}, \texttt{[Low1/64]BRank(12):256(2)}, \texttt{[Low1/64]BRank(12):384(1)} \\
\texttt{xoroshiro128aox-24-16-37} & 77  & 45500 & 32 TB  & \\
\texttt{xoroshiro128aox-55-14-36} & 66  & 45500 & 32 TB  & \\
\hline
\end{tabular}
\label{practrand-results}
\end{table*}

\subsection{Gjrand}


Table \ref{gjrand-results} lists the results for the Gjrand test set. The
number of test failures is the total number of tests across all 100 seeds that
exhibit $p$-values outside of the range $[0.001, 0.999]$. All variants of
\texttt{xoroshiro128} fail systematically on both versions of the \texttt{z9}
Hamming-Weight dependency test.
The Mersenne Twister fails Binary Rank (\texttt{binr}) systematically.

\begin{table}[!t]
\renewcommand{\arraystretch}{1.3}
\centering
\caption{Results for the Gjrand test set}
\begin{tabular}{lcl}
\hline
Generator & Test failures & Systematic failures \\
\hline
\texttt{mt32}                     & 107 & \texttt{binr -c} \\
\texttt{pcg64}                    & 15  & \\
\texttt{philox4x32-10}            & 7   & \\
\texttt{xoroshiro128+-24-16-37}   & 257 & \texttt{z9}, \texttt{z9 -t} \\
\texttt{xoroshiro128+-55-14-36}   & 286 & \texttt{z9}, \texttt{z9 -t} \\
\texttt{xoroshiro128aox-24-16-37} & 210 & \texttt{z9}, \texttt{z9 -t} \\
\texttt{xoroshiro128aox-55-14-36} & 205 & \texttt{z9}, \texttt{z9 -t} \\
\hline
\end{tabular}
\label{gjrand-results}
\end{table}

\subsection{Hamming Weight Dependency test}

Table~\ref{hwd-results} lists the results for the HWD tests. Each generator is
run until it generates a $p$-value smaller than $10^{-20}$ or outputs 100~TB of
data.
Due to the long test runtime, results are given for a single 128-bit seed
(\texttt{s0} = 1, \texttt{s1} = $-1$). We choose this more extreme $p$-value
bound to be consistent with the results published for other generators,
e.g.~\cite{Blackman21}, but also list the data output for $p=10^{-3}$ for
comparison with similar tests using our $[0.001, 0.999]$ $p$-value failure
range.

In both shift variants of AOX, biases in the output take substantially more
output to detect. The $p=10^{-20}$ results are comparable to the
\texttt{xoshiro256} (54~TB) and \texttt{xoroshiro1024+} (36~TB), generators
with twice and eight times more state than \texttt{xoroshiro128aox}
respectively.
The results for $p=10^{-3}$ are comparable with the Gjrand \texttt{z9} test
failures for \texttt{xoroshiro128} generators within 10~TB.

Although the Mersenne Twister is an \ftwo-linear generator, it is unsurprising
that the HWD test does not detect anomalies because it has a significantly
larger state and would require a correspondingly huge amount of memory to do
so. When reduced to 607 bits of state, the Mersenne Twister fails HWD at 37~GB
of output~\cite{Blackman21}.

\texttt{pcg64} and \texttt{philox4x32-10} do not exhibit any bias up to 100 TB
of output, but neither are based on \ftwo-linear maps.

\begin{table}[!t]
\renewcommand{\arraystretch}{1.3}
\centering
\caption{Data output from a generator to reach a particular $p$-value threshold
  under the 64-bit HWD test.}
  \begin{tabular}{lp{1.8cm}p{1.8cm}}
\hline
Generator &
$p=10^{-3}$ &
$p=10^{-20}$ \\
\hline
\texttt{mt32}                     & $>$100 TB & $>$100 TB \\
\texttt{pcg64}                    & $>$100 TB & $>$100 TB \\
\texttt{philox4x32-10}            & $>$100 TB & $>$100 TB \\
\texttt{xoroshiro128+-24-16-37}   & 1.8 GB & 4 TB \\
\texttt{xoroshiro128+-55-14-36}   & 1.1 GB & 4 TB \\
\texttt{xoroshiro128aox-24-16-37} & 1.8 TB & 36 TB \\
\texttt{xoroshiro128aox-55-14-36} & 11.4 TB & 45 TB \\
\hline
\end{tabular}
\label{hwd-results}
\end{table}

\subsection{Bitwise linear complexity}

Each bit of the output of the \texttt{xoroshiro128aox} (55-14-36) is tested for
linear artefacts with TestU01's \texttt{smarsa\_MatrixRank} with a binary matrix of
size $10,000 \times 10,000$, four times the size of the largest matrix used in
BigCrush and \texttt{scomp\_LinearComp} test with a sequence length of 800,000,
double that of the longest used by BigCrush.
Each bit is tested over 100 seeds as per the methodology described in
Section~\ref{methodology}.
No systematic failures were found for any of the 64 output bits in either test.

\section{Hardware implementation cost}
\label{implementation-cost}

To assess the cost of the \texttt{xoroshiro128aox} generator in hardware, we
compare it against the contemporary generators included in the statistical
testing.
To do this, we have produced Verilog RTL (register transfer level)
implementations of the generators and performed synthesis and physical place
and routing using Synopsys ASIC tooling. We use Graphcore's 7~nm cell library
and a target clock period of 1~GHz. In each implementation, a generator
computes its state update and output function in a single cycle, reading and
writing to and from registers within the block. Reported gate counts only
include combinatorial logic associated with these functions.
We omit \texttt{mt32} from this implementation analysis since the hardware cost
of its state alone (19,937 bits) is prohibitively expensive.

\begin{figure}[!t]
  \begin{center}
  \includegraphics[scale=0.22]{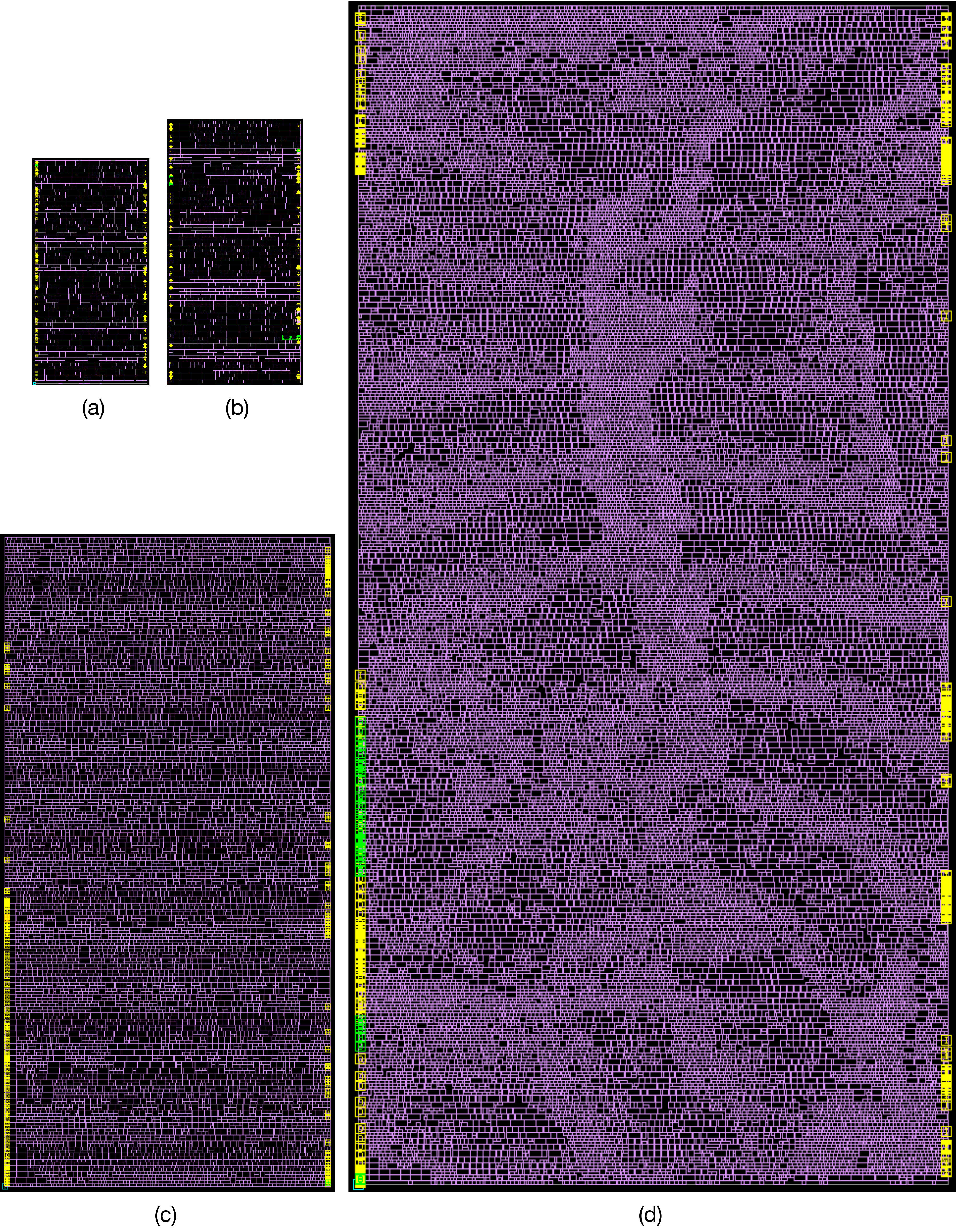}
  \end{center}
  \caption{Comparison of physical floorplans for (a) \texttt{xoroshiro128aox}, (b) \texttt{xoroshiro128plus}, (c) \texttt{pcg64} and (d) \texttt{philox4x32-10}.}
  \label{floorplans}
\end{figure}

The hardware implementation costs of the generators is measured by the number
of gates required and the logical depth, which is the maximum number of gates
of any path through the logic function. The results of the analysis are
summarised in Table~\ref{gen-hw-costs} and are presented for the state update
state update and the output function. Figure~\ref{floorplans} shows the scaled
floorplans of each generator for comparison, including the state-transition
logic, output function and state.
The \texttt{xoroshiro128} function can be implemented with three 64-bit XORs.
The shift and rotate operations are by constant values and require no logic.
The cost of the AOX output function is similar to the state update, and the
full variable 64-bit addition is approximately three times as expensive as AOX.

For \texttt{pcg64}, the main hardware components are a 128-bit constant
multiplier and a 128-bit constant adder for the state update, which dominates
the hardware cost with $\sim$10,000 gates and 27 levels of logic.
Note that there is significant scope for optimising the implementation of
constant multipliers since partial products will be generated for each set bit
in the constant. Assuming approximately 50\% set bits in the constant, only
$n/2$ partial products need to be accumulated.
The output function requires a 64-bit XOR of the state vector and 64-bit full
barrel rotator, which is cheap to implement compared with a full 64-bit adder.

\texttt{philox4x32-10} is the most expensive generator to implement due to its
complex output function consisting of 10 stages of four 32-bit constant
multipliers, two 32-bit constant adders and four 32-bit XORs of the \texttt{k}
and \texttt{l} values. The implementation cost is $\sim$30,000 cells and 89 stages
of logic. In any practical implementation, the output function would need to be
heavily pipelined to meet an acceptable clock speed.
Fewer rounds could also be implemented if the quality of the output were
satisfactory, however we have not investigated that trade off.
The state transition as expected is relatively cheaper, requiring a 128-bit
increment by one.

\begin{table*}[!t]
\renewcommand{\arraystretch}{1.3}
\centering
\caption{Hardware costs of different PRNGs as measured by gate counts}
\begin{tabular}{lccccc}
\hline
  & \multicolumn{2}{c}{State update} & \multicolumn{2}{c}{Output function} & \\
  Generator & Total cells & Logic depth & Total cells & Logic depth & Total cells \\
\hline
\texttt{xoroshiro128aox}  & 331   & 4    & 353    & 4  & 684 \\
\texttt{xoroshiro128plus} & 331   & 3    & 906    & 13 & 1,237 \\
\texttt{pcg64}            & 9,564 & 26   & 658    & 7  & 10,222 \\
\texttt{philox4x32-10}    & 1,003 & 13   & 29,553 & 89 & 30,556 \\
\hline
\end{tabular}
\label{gen-hw-costs}
\end{table*}


\section{Other aspects of quality}
\label{other-quality}

Apart from statistical quality of the output of a PRNG, several other issues
affect the suitability for use in artificial intelligence applications. This
section discusses these: period, output uniformity, seeding and overlapping
parallel sequences.


\subsection{Period}

The period of a PRNG is the number of states that are visited before the
sequence of states repeats. Since \texttt{xoroshiro128} has a period of
$2^{128}-1$ (excluding the all-zeros state from which it cannot transition)
\texttt{xoroshiro128aox} has the same period. This period is sufficient to
accommodate many parallel generators, as is discussed in
Section~\ref{parallel-generators}. To give an indication of how large this
period is: if a single generator were to output a value every nanosecond (1
billion times a second), it would take $10^{22}$ years to traverse the whole
sequence.

\subsection{Uniformity}
\label{uniformity}



Uniformity of a PRNG is a measure of how evenly distributed the different
output values are. A perfectly uniform generator will output all distinct
values an equal number of times after completing a full period, whereas a
non-uniform generator biases particular values. Since the \texttt{xoroshiro128}
PRNG is full period (notwithstanding the all-zeros state), an analysis of
uniformity can be focused on the AOX output function.


A small non uniformity is evident when directly calculating the distribution of
output values for outputs for small AOX output sizes.
To measure this non uniformity, we use the $\chi^2$ test for a discrete uniform
distribution as a goodness-of-fit test when compared with the $\chi^2$
distribution.
Since it is intractable to measure all or even a large part of the full 128-bit
state space, the $\chi^2$ statistic is calculated for smaller states up to 40
bits and all possible output values, and extrapolate the result to a 128-bit
state size.
AOX is a function that maps $2n$-bit state values to $n$-bit outputs. If we
take $m$ samples of AOX, then the expected number of occurrences for any output
value is $\frac{m}{2^n}$, according to the null hypothesis.
This test for uniformity is similar to running TestU01's
\texttt{smultin\_MultinomialBits} as a collision test.

Calculating the test statistics up to 40 state bits and 20 output bits and
comparing them to the critical values at a 95\% significance level (for 20
state bits $\chi^2=373,621$ and the critical value is 1,050,430) shows that
there is no statistically significant difference between the output of AOX and
the uniform distribution for any sample sizes. This strongly suggests that the
non uniformity cannot be detected through sampling of the 128-bit state space.

\subsection{Escaping zero land}

A desirable property of a PRNG is that given a 'bad' state where only a
minority of bits are set to one, it can rapidly transition to a 'good' state
where approximately half the bits are set, such as from a poor initial seed, or
a bad state it encountered in its sequence. This capability is often referred
to as \emph{escaping zero land}, and for linear generators is equivalent to the
ability for correlated states to decorrelate quickly. In general, zero escape
and decorrelation is a problem for generators with a large state space, where
the transition function must spend more time perturbing the state, at the cost
of performance/implementation cost.

This issue is relevant to PRNG initialisation since seed values are typically
not uniform random bits. In intelligence applications, deterministic execution
is important for debugability and so fixed seed values are required. Efficient
generation of fixed seeds is important to avoid memory use and to minimise
initialisation overheads, but this makes it difficult to ensure they are good
values. A similar issue arises when a generator encounters a bad state in its
sequence of transitions. In this case, it should recover quickly back to
well-balanced states.

\begin{figure*}[!t]
\begin{minipage}[b]{0.5\linewidth}
\centering
\includegraphics[width=3.13in]{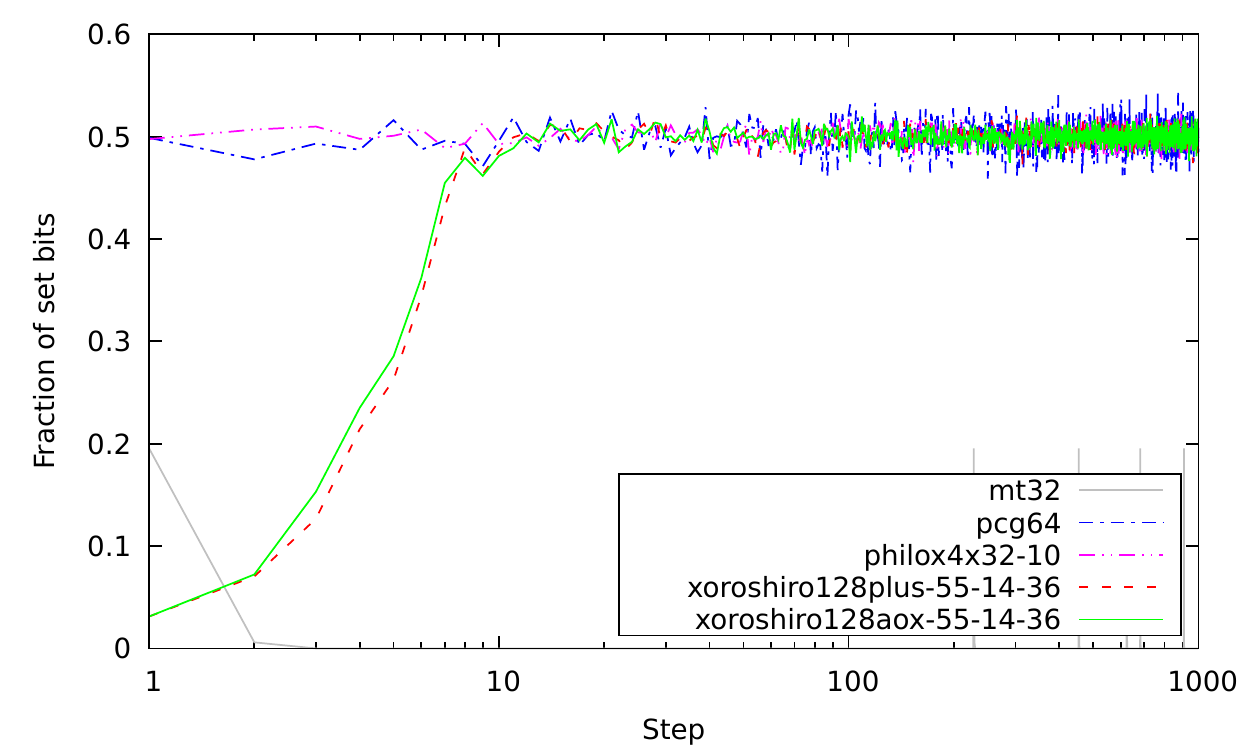}
\caption{Convergence to half of the output bits being set (first 1K samples).}
\label{zero-escape-1k-fig}
\end{minipage}
\hfill
\begin{minipage}[b]{0.5\linewidth}
\centering
\includegraphics[width=3.13in]{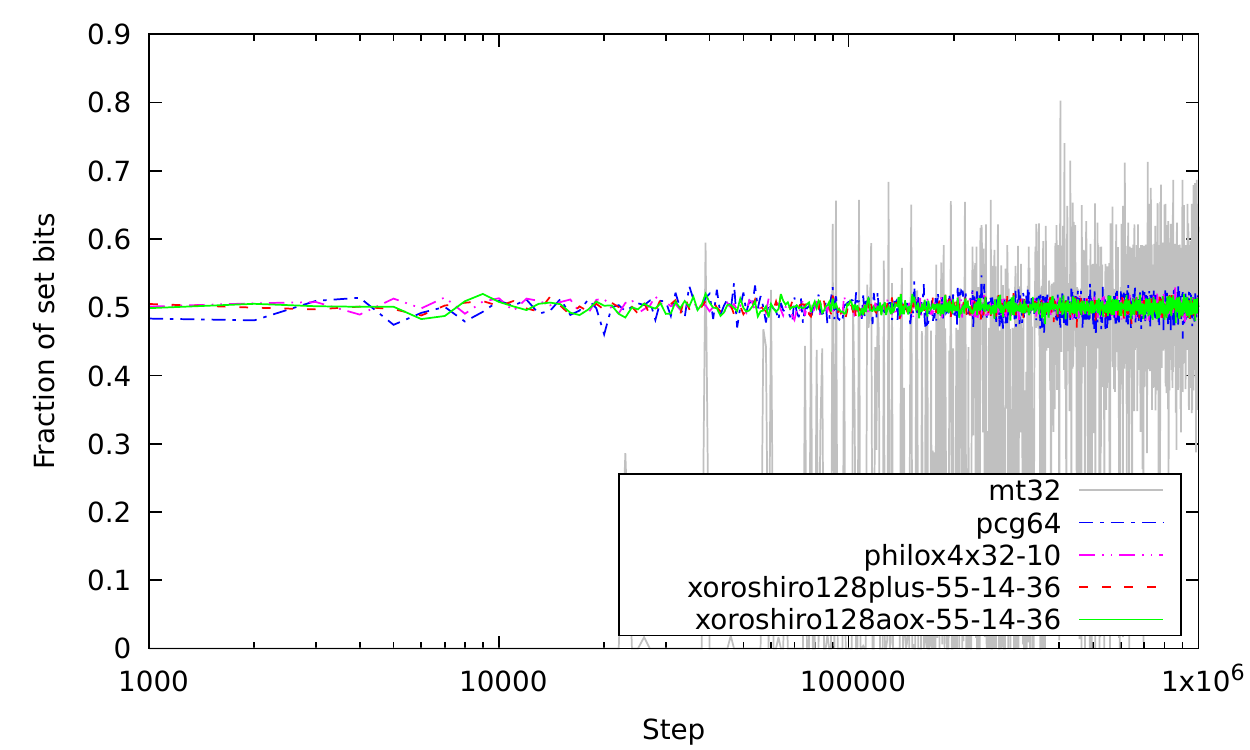}
\caption{Convergence to half of the output bits being set (1K to 1m samples).}
\label{zero-escape-1m-fig}
\end{minipage}
\end{figure*}

To characterise the rate at which a generator escapes from zero land, we use
the method of Panneton, L'Ecuyer and Matsumoto~\cite{Panneton06}. A
generator is initialised with a one-hot seed, and the proportion of set bits in
the output is recorded over a fixed number of generated values, averaged over
the last four outputs. The escape time is calculated by averaging the
proportion at each output over all one-hot seeds. The results of this analysis
are shown in Figure~\ref{zero-escape-1k-fig} (1,000 iterations) and
Figure~\ref{zero-escape-1m-fig} (1 million iterations, sampled at intervals of
1,000).

\texttt{pcg64} and \texttt{philox4x32-10} produce balanced outputs immediately.
For \texttt{philox4x32-10} it is necessary for the output to be balanced
regardless of the state since it is advanced as a counter. The AOX output
scrambler has a very similar behaviour to addition, with escape time being
approximately 12 iterations, relating mainly to the ability of the
\texttt{xoroshiro128} PRNG transition function to decorrelate. The
\texttt{mt32} generator takes over a million cycles to reach an approximately
balanced output state, due to it having a much larger state.\footnote{It should
also be noted that the \texttt{mt32} state does not begin one hot: the Boost
C++ implementation used in these experiments performs an operation on the state
after it is set directly to improve the balance of set bits. Without this
operation, the warmup period would be further extended.}

\subsection{Parallel generators}
\label{parallel-generators}

Since the IPU's \texttt{xoroshiro128aox} generator is used in the context of
large amounts of parallelism, with each chip containing more than a thousand
processing tiles and a system deployment containing many IPUs, it is pertinent
to ask whether 128 bits of state is sufficient, particularly since the authors
of \texttt{xoroshiro128+} recommend its state space is large enough only for
mild parallelism.\footnote{A comment in the source code for
\texttt{xoroshiro128+} at \url{https://prng.di.unimi.it/xoroshiro128plus.c}.}
When producing parallel outputs from a generator, two issues of concern are
whether sequences will overlap and consequently be correlated, and whether
non-overlapping sequences are correlated.

To address the first question, since a jump function~\cite{Haramoto08} can be
used to initialise \texttt{xoroshiro128} seeds at specific offsets in the
sequence and therefore guarantee they are non overlapping. For example, a jump
function can be used to produce $2^{64}$ unique sequences of length $2^{64}$
(or 128 exabytes of data).
Even if seeds were to be chosen randomly within the sequence, the probability
of overlap is very small. Using an upper bound on this
probability~\cite{Vigna20}, if $n$ is the number of generators, $L$ is the
sequence length and $P$ is the period length, then the probability of overlap
is at most $n^2 L/P$. This bound assumes the generator is full period in that
there is a single cyclic sequence of transitions between states, which is true
for the \texttt{xoroshiro128} PRNG when excluding the zero state.
In an extreme scenario, with a Graphcore machine containing 65,536 IPU
processors and approximately 0.5~billion parallel generators,\footnote{Based on
a total of 1,216 tiles per chip, 6 hardware threads per tile, a 1~GHz clock
speed with a thread running at 1/6~GHz.} performing two state updates every
cycle and running for 32 days, the probability of two sequences overlapping is
negligible at 0.00006\%.

To address the second question of correlated non-overlapping sequences, we
perform a set of additional tests on \emph{interleaved generators}, which
produce output in a round-robin manner from $N$ independent generators.
We take a straightforward approach by choosing interleave factor of 1 and $N=$
$10$, $100$ and $1000$. Each generator is tested using PractRand up to 32~TB of
output, or 4 billion outputs when $N=1000$ (much more than TestU01's BigCrush
can consume).
The above two seeding schemes are used: unique sequences of length $2^{64}$ and
randomised start points. All six tests run to completion without any failures
being flagged.

\section{Conclusions}
\label{conclusion}

In this paper we have presented the IPU's PRNG algorithm, the AOX variant of
\texttt{xoroshiro128}, and provided a rigorous assessment of its statistical
quality. Our analysis goes well beyond the typical testing of PRNGs found in
the literature, which often only present results for a particular test suite
and sometimes only for a single seed. We provide results for all well-regarded
test suites that we are aware of, as well as adopting the approach of sampling
generators to test over 100 different seeds. For Test U01's BigCrush, which is
widely regarded as the standard for PRNG testing, we take the additional step
of testing each generator with various permutations of the output bits to
mitigate biasing particular bits and to expose weaknesses in the
\texttt{xoroshiro} family of generators, particularly with addition as an
output function.

Our results show that the AOX variant of \texttt{xoroshiro128} passes BigCrush
under all output types and passes PractRand's standard tests up to 32~TB of
output, where in both cases \texttt{xoroshiro128+} demonstrated systematic
failures. These results indicate that the weakness of the addition output
function have largely been mitigated by AOX. However, PRNGs based on
\ftwo-linear maps are known to exhibit linear artefacts that can always be
detected given analysis of enough output, or by a particular test. A scrambling
of the PRNG's output can only serve to hide the linearities to an extent. As
such, the \texttt{z9} test of Gjrand and the related HWD test both detect
dependencies in the populations of set bits between consecutive outputs for
addition and AOX \texttt{xoroshiro128} variants.
No such dependencies are detected by similar tests in BigCrush or PractRand, or
by focused testing of each bit.

Although AOX is not perfectly uniform, and were it to be implemented in
software, less performant than integer addition, the \texttt{xoroshiro128aox}
PRNG represents a significant improvement in statistical quality over
\texttt{xoroshiro128+} passing both major test suites BigCrush and PractRand,
while being cheaper to implement in hardware and suitable for inclusion in a
processor that is instanced thousands of times on a single chip.
Contemporary fast PRNGs with the same state size, \texttt{pcg64} and
\texttt{philox4x32-10}, are more robust to tests for linear artefacts, but are
orders of magnitude more expensive to implement in hardware and thus
prohibitively expensive for use in the IPU's tile processors.
\texttt{xoroshiro128aox} therefore provides a good trade off between
implementation cost and statistical quality.

\ifCLASSOPTIONcaptionsoff
  \newpage
\fi



\bibliographystyle{IEEEtran}
\bibliography{IEEEabrv,refs.bib}
\end{document}